\documentclass[traditabstract]{aa}

\usepackage{natbib}
\usepackage{graphicx}
\usepackage{txfonts}
\usepackage[bookmarks,bookmarksnumbered,colorlinks=true, citecolor=blue,anchorcolor=blue, urlcolor=cyan, breaklinks=true, menucolor=black, linktocpage=true]{hyperref}

\def\arcsec{\hbox{$^{\prime\prime}$}}
\def\kms{{${\rm km}\:{\rm s}^{-1}\:\!\!$}}
\def\kmss{{${\rm km}\:{\rm s}^{-2}\:\!\!$}}

\def\deg{\mbox{$^{\circ}$}}

\begin{document}

   \title{Supergranular-scale magnetic flux emergence beneath an unstable filament}
 
\titlerunning{Supergranular-scale flux emergence beneath an unstable filament}

   \subtitle{}

 \author{J.~Palacios\inst{1}, 
 C.~Cid\inst{1}, A.~Guerrero\inst{1}, E.~Saiz\inst{1}, \and Y.~Cerrato\inst{1}  }

   \institute{$^{1}$Space Research Group - Space Weather, Departamento de F\'isica y Matem\'aticas, Universidad de Alcal\'a, Spain\\
              \email{judith.palacios@uah.es}
             }
\authorrunning{J.~Palacios et al.}

   \date{Received ---------; accepted -------------}

 
  \abstract
   {}
   {Here we report evidence of a large solar filament eruption on 2013, September 29. This smooth eruption, which passed without any previous flare, formed after a two-ribbon flare and a coronal mass ejection towards Earth. The coronal mass ejection generated a moderate geomagnetic storm on 2013, October 2 with very serious localized effects. The whole event passed unnoticed to flare-warning systems. }
   {We have conducted multi-wavelength analyses of the Solar Dynamics Observatory through Atmospheric Imaging Assembly\ (AIA) and Helioseismic and Magnetic Imager (HMI) data. The AIA data on 304, 193, 211,~and 94 \AA~sample the transition region and the corona, respectively, while HMI provides photospheric magnetograms, continuum, and linear polarization data, in addition to the fully inverted data provided by HMI.}
   {This flux emergence happened very close to a filament barb that was very active in mass motion, as seen in 304~\AA~images. The observed flux emergence exhibited hectogauss values. The flux emergence extent appeared just beneath the filament, and the filament rose during the following hours. The emergence acquired a size of 33\arcsec~in $\sim$12~h, about $\sim$ 0.16 \kms. The rate of signed magnetic flux is around 2$\times$10$^{17}$~Mx min$^{-1}$ for each polarity. We have also studied the eruption speed, size, and dynamics. The mean velocity of the rising filament during the $\sim$ 40 min previous to the flare is 115$\pm$5~\kms, and the subsequent acceleration in this period is 0.049$\pm$0.001~\kmss.}
   {We have observed a supergranular-sized emergence close to a large filament in the boundary of the active region NOAA11850. Filament dynamics and magnetogram results suggest that the magnetic flux emergence takes place in the photospheric level below the filament. Reconnection occurs underneath the filament between the dipped lines that support the filament and the supergranular emergence. The very smooth ascent is probably caused by this emergence and torus instability may play a fundamental role, which is helped by the emergence.}


   \maketitle
%

\section{Introduction}

On 2013, September 29 a large solar filament rose smoothly, without any previous flare, vigorous dynamics, or significant brightening arcade. This filament lifted and afterwards created a two-ribbon flare,  and its X-ray flux reached the category C1.2 by GOES 1-8~\AA\footnote{\url{http://www.lmsal.com/solarsoft/last_events_20131001_2320/index.html}}. The eruption led to an Earth-directed halo coronal mass ejection (CME). The interplanetary plasma interacted with the Earth's magnetic field creating a moderate geomagnetic storm with effects on the ground \citep[{\it Dst}  geomagnetic index reached --75 nT,  and storm classification follows ][]{Gonzalez1994} on 2013 October 2\footnote{{\it Dst} data available in \url{http://wdc.kugi.kyoto-u.ac.jp/dst_provisional/201310/index.html}}. However, since the interplanetary medium and terrestrial effects are beyond the scope of this paper, they will be treated elsewhere. This eruption went unnoticed for flare early alerts and predictions and it produced a geomagnetic storm. This case is a very pertinent example, which  can be used to obtain input for simulations and for solar and space weather predictions and warnings. 

In this paper, we  focus on the solar source and trigger of this filament eruption, which seems to be a supergranular-sized magnetic flux emergence occurrence underneath the filament. We investigate  the reasons for this  kind of an undisturbed rise using Solar Dynamics Observatory (SDO) data by means of the Atmospheric Imaging Assembly (AIA) at various EUV wavelengths (AIA/SDO 304~\AA, 193~\AA, 211\AA~and 94~\AA). Also, to explain the event we obtained a close-up using Helioseismic and Magnetic Imager (HMI) magnetograms, continuum, and linear polarization images. We also took advantage of the fully inverted HMI data in this work.

In the review of \citet{Linker2003} it is stated that only 20\% of the CMEs are associated with a large flare. Flareless CMEs have been reported by \citet{Song2013}. \citet{Gosain2012} state that most of the filament eruptions related to CMEs pass undetected because of their low emission and, consequently, are not associated with flares. Therefore, trusting only in flare occurrence may be misleading and the region can actually be CME-productive. 

Some very recent and comprehensive summaries of the various models that can trigger the eruption of filaments and CMEs are \citet{Aulanier2010, Schmieder2013, Aulanier2014, Parenti2014} and references therein; the "magnetic breakout model" \citep{Antiochos1998, Antiochos1999}; a highly twisted flux rope, as in the "torus instability" \citep{Torok2003, Torok2005, Kliem2006}; different instabilities, e.g. MHD \citep{Klimchuk1989, Sturrock1989}; catastrophic loss of equilibrium \citep{Forbes1991};  approaching polarities \citep{Forbes1995}; moving features and magnetic cancellation \citep{Ballegooijen1989, Forbes1991}; and the "tether cutting"  model, implying reconnection in the arcades below the flux rope \citep{Moore1992, Moore2001}. Considering the magnetic flux emergence,  filament eruptions due to flux emergence in the surroundings have been analysed in simulations by \citet{Chen2000}. The magnetic configuration of active and quiescent filaments and their CME productivity have been studied in \citet{Feynman1995}. Slow and fast eruptive phases have been studied by \citet{Sterling2005, Sterling2007}.
 We  consider the observational features of these models to choose the most plausible model for this phenomenon.
  
More particularly, different works are revised in the review by \citet{Chen2011}: \citet{Zhang2008} studied events of flux emergence and cancellation 12 h prior to a CME initiation, finding that 91\% are related. \citet{Wang1999} also found filament eruptions close to flux emergence regions. \citet{Jing2004} detected that more than half of the eruptive filaments were related to flux emergence. \citet{Zhang2001} related relatively small flux cancellation areas with the filament eruption of the Bastille Day event. 

However, many of these works were qualitative, only with the presence and closeness of the flux emergence to relate with other magnitudes. Here we aim to describe the filament quantitatively and to assess its instabilities, and we describe how a flux emergence can influence a filament lift-off. 

This paper is organized as follows: first, the observations are described in Section~\ref{s1}, describing the flux emergence and filament dynamics, along with the surrounding coronal holes. All calculations are included in Section~\ref{s1}. The discussion and conclusions are presented in Sections~\ref{s2} and \ref{s3}, respectively.

\section{Observations of the emerging region and filament lift-off}\label{s1}

We used data provided by the space solar facility Solar Dynamics Observatory \citep[SDO, ][]{Pesnell2012}. Data were acquired with the Atmospheric Imaging Assembly \citep[AIA, ][]{Lemen2012} and the Helioseismic and Magnetic Imager \citep[HMI, ][]{Scherrer2012}.  The AIA multi-wavelength data comprise observations in 304 and 193~\AA, which sample the high chromosphere, transition region, and corona, respectively. We  also included AIA 211~\AA~data, which are useful as an intermediate wavelength measurement since the filament opacity, compared to 304~\AA, is somewhat reduced. The AIA 94~\AA~data also probe the inner corona. The HMI data sample the Fe $\textsc{i}$ line at 6173~\AA~to generate line-of-sight (hereafter, LOS) photospheric magnetograms. This line samples a photospheric height around $\sim$100~km \citep{Fleck2011}. The data cadence for the general study is chosen to be 12~min from September 29 to October 1 on both AIA and HMI instruments. In  other particular cases of the study, the cadence is reduced to 1~min or increased up to 180~min (3~h). Images are co-spatial and quasi-synchronized (within seconds). Pixel sampling for both instruments is 0\arcsec6.

All images are reduced through the SolarSoftWare (SSW)  $\it{read\_sdo.pro}$ and $\it{aia\_prep.pro}$ routines. Next, frames are corrected from rotation and aligned. 

\begin{figure*}
\begin{center}
\includegraphics[width=0.52 \linewidth]{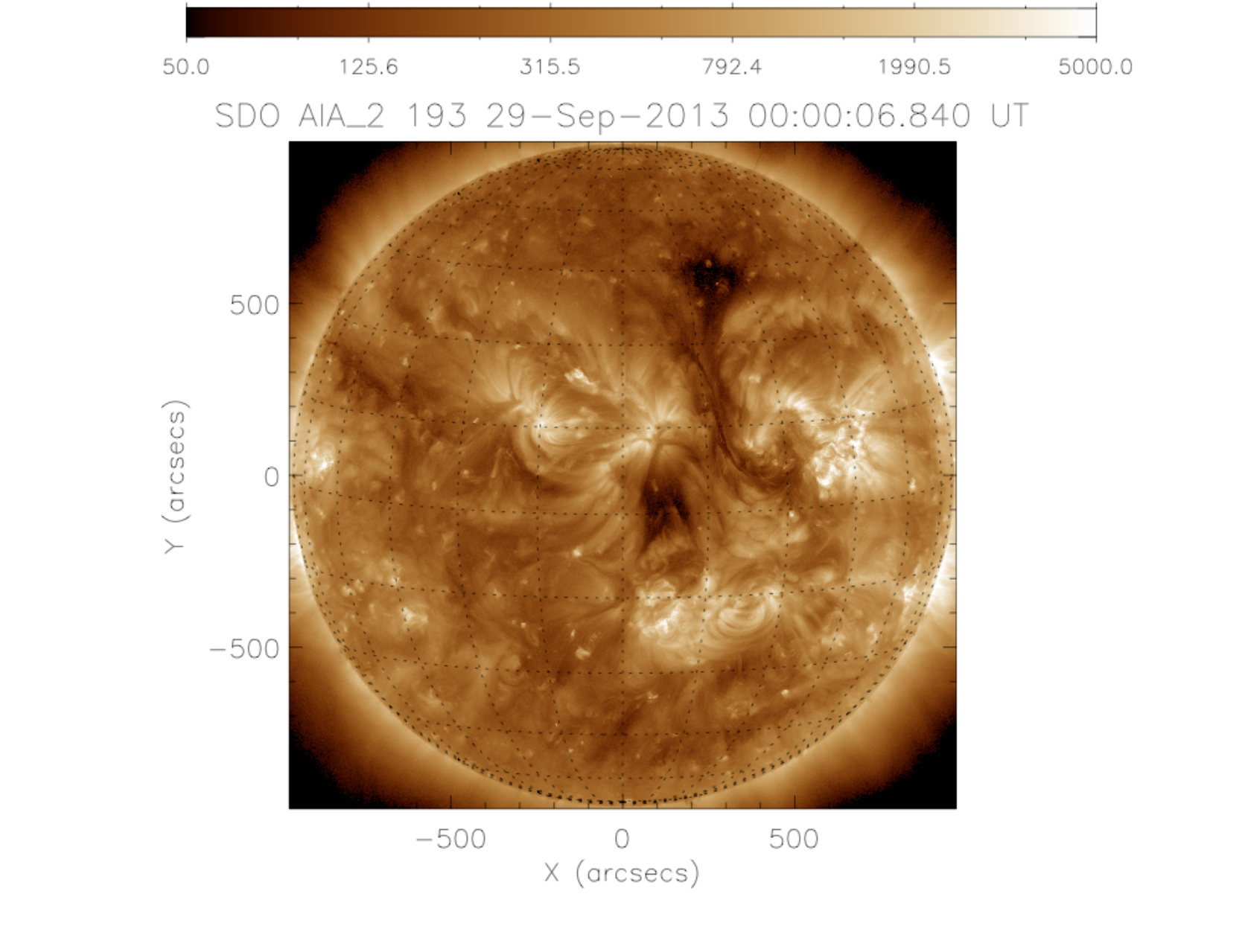}~~
\includegraphics[width=0.52 \linewidth]{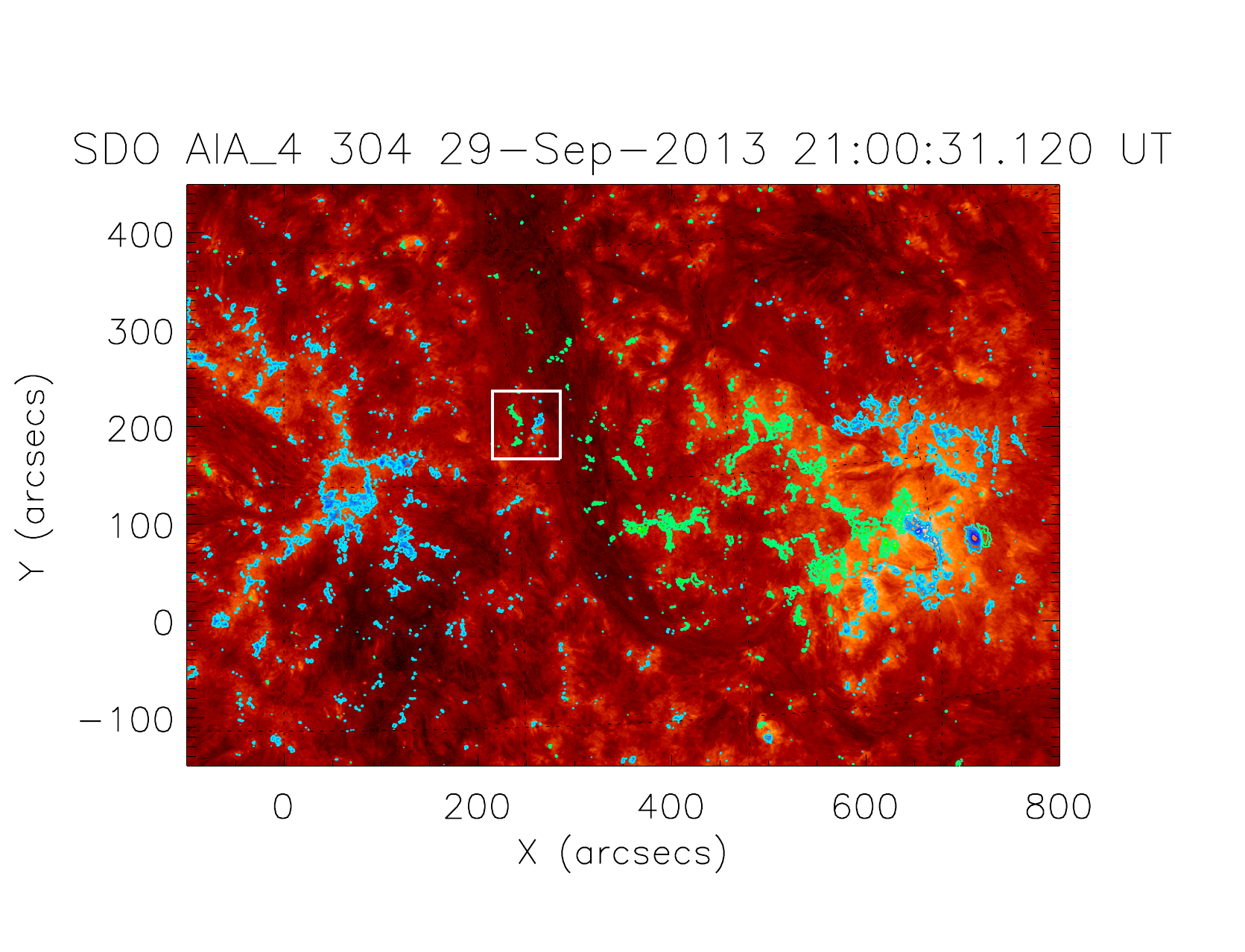}

\caption{Context figures of the filament. The grid spaces mark 15\deg. {\it Left:} Image in AIA 193~\AA, on 2013 Sept 29, at 00:00~UT. Image is clipped from 50 to 5000 DN. The filament is located on 20$\deg$N 15$\deg$W, with two coronal holes surrounding it on coordinates 40$\deg$N 20$\deg$W (CH1) and 0$\deg$N 7$\deg$W (CH2).  
{\it Right:} Context image showing a derotated AIA 304~\AA~image displaying the filament, and the positive polarity (in green) and negative (blue). Different colour shades are explained in caption of Fig.~\ref{fig2}. The bipolar flux emergence occurrence is located at [250\arcsec, 200\arcsec], enclosed in a white square. A movie available in the online data base shows the temporal evolution of the whole visible hemisphere from 21:00 to 22:09 UT in the AIA 304~\AA~channel. In the movie, the FOV of the right panel of this figure is marked by a black rectangle, and the emergence by a white square. The green and blue contours have the same meaning as in the right panel.}
\label{fig1}
\end{center}
\end{figure*}

Figure~\ref{fig1} shows context images. The left panel indicates the whole solar hemisphere, showing the filament and two surrounding coronal holes in AIA 193~\AA. The right panel shows the filament in AIA 304~\AA~as a dark, inverse-S-like, elongated structure located between two opposite polarities of  NOAA AR11850 and the following active region. The filament was visible close to the trailing boundary of AR 11850, with its southern part close to its positive facular polarity, and the spine located at the polarity inversion line (PIL), with a facular negative polarity on the left and another positive on the right. As seen in AIA 304 and 193~\AA, in close to five days prior to the eruption, the filament developed from a short structure (Sept 25, 21:00~UT) to a inverse-S-shaped elongated in just one day. This fully developed filament evolved in a dark area in 193~\AA, which is an open field region. The length was about $\sim$0.40$R_{\sun}$ on Sept 25 to $\sim$0.63$R_{\sun}$ two days later. This filament showed very active mass motion along the spine and  barb. The southern part is more elevated than the northern part at this stage. The heights look similar at Sept 28, 10:00~UT. This situation seems to reverse on Sept 28 19:00~UT. The magnetic flux emergence started around Sept 29 at 00:00~UT and the appearance of pushing the filament from below is conspicuous. This emergence displayed small pores in continuum. There was a positive polarity where one of the barbs was located, and this region seems to be attaching the filament to the surface. The apex of the filament rose, placed somewhat northern from the barb, around 21:00~UT. Eventually the filament developed a two-ribbon flare on Sept 29 at 21:43~UT, which led to a maximum flux of X-ray of C1.2 in GOES. The southern leg was detached early in the flare, and the northern leg produced a shower of coronal rain on Sept 29, at 23:29~UT, lasting around one hour. Thread-like features of the filament feet appeared to fall down. 

The subsequent CME appeared in LASCO at 22:24~UT, with a plane-of-the-sky speed (POS) of $\sim$ 600 \kms . The halo CME is reported in the CACTUS catalog \citep[\url{http://sidc.oma.be/cactus/catalog/LASCO/2_5_0/qkl/2013/09/latestCMEs.html}, ][]{Robbrecht2004, Robbrecht2009}. However, the reported speed in the LASCO catalog (\url{http://cdaw.gsfc.nasa.gov/CME_list/}) is 1179 \kms. This discrepancy is due to the external CME shell, which is faster than the filament.

\subsection{Magnetic flux emergence}

\begin{figure*}
\begin{center}
\includegraphics[width=1.0\linewidth, angle=0.]{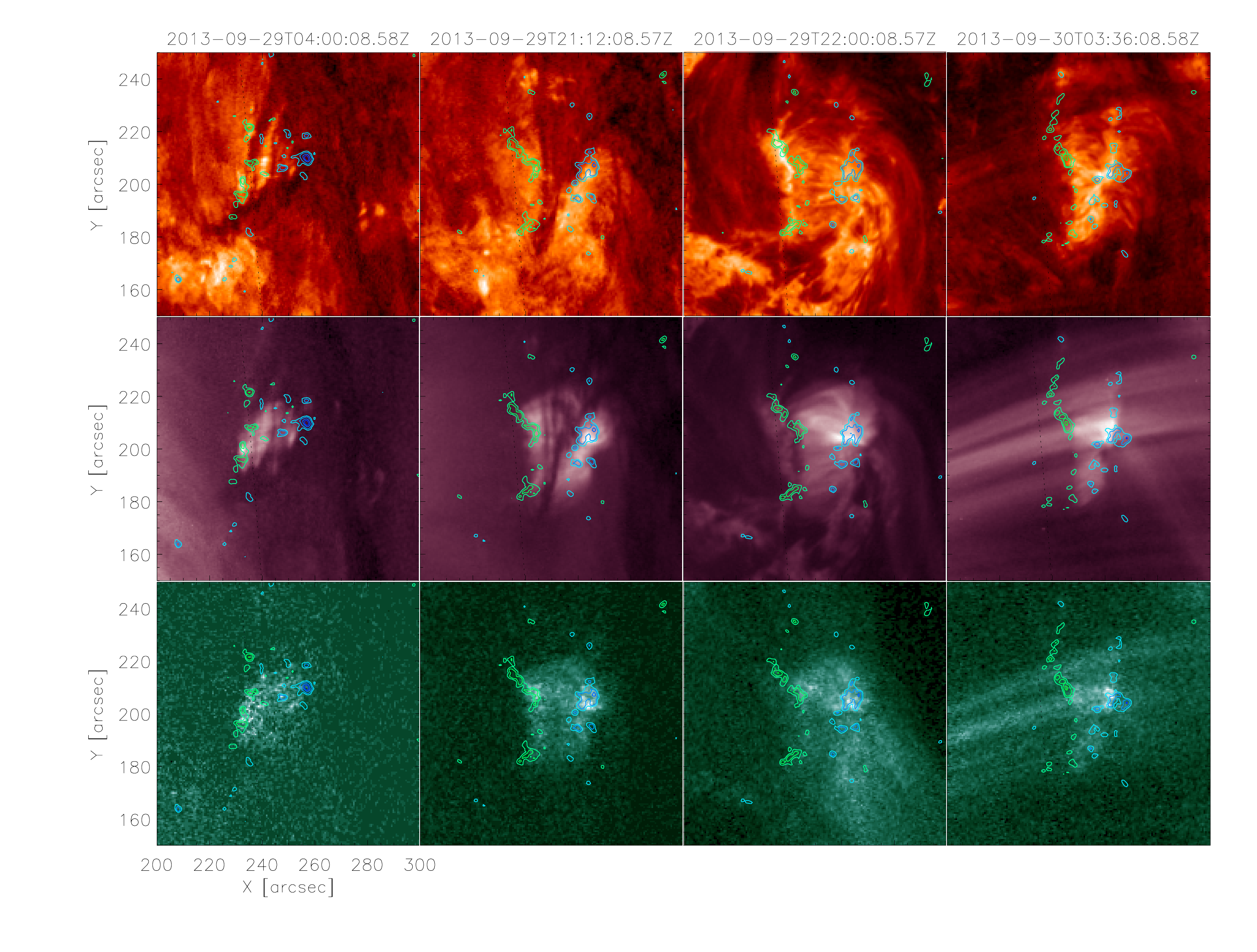}
\caption{{\it Top row}: frame sequence of AIA 304~\AA~images of a filament's barb {\it (darker~areas)} with LOS-magnetic field superimposed. Different shades of green (from lighter to darker) contours mark different levels, as 90, 200, 500, 600, and 700 G, while a gamut of blue contours mark -90, -200, -500, -700, -900 G for the negative polarities.\ The frame size is 100\arcsec $\times$ 100\arcsec. Sequence timing aims to show the magnetic flux emergence (panel 1), pre-flare (panel 2), flare (panel 3), and decay (panel 4), from left to right. {\it Bottom row}: same temporal sequence in AIA 211 and 94~\AA. Movie issued as on-line material. }
\label{fig2}
\end{center}
\end{figure*}

We  selected AIA 304~\AA~images and HMI magnetograms with a cadence of 12 min, from Sept 29 00:00~UT to Oct 1 10:00~UT. We  analysed  AIA 304~\AA~close-up images with overlaid HMI magnetograms of the aforementioned filament barb (located on $\sim$ 18\deg N, 15\deg W in the left panel of Fig.~\ref{fig1}), which appears to be the location where the photospheric magnetic flux emergence happens. The panels in Fig.~\ref{fig2} show the evolution, displaying selected frames\footnote{ The temporal evolution shown in Fig.~\ref{fig2} is available in the on-line edition.}. This barb seems to be close to a positive polarity, on Sept 29 at 00:00~UT. A large magnetic supergranular extent appeared, as seen in the HMI magnetograms. This large supergranular emergence exhibited mixed polarities at first, and later on the positive and negative polarities drifted, rearranging as ordinary  network elements. Three hours later, the main negative polarity started developing a pore. The positive polarities also started developing small pores  in the following hours. Six hours later, this emergence extent, consisting of positive polarities {\it (green)} on the left and negative polarities {\it (blue)} on the right, moved apart up to $\sim$ 25\arcsec\ (panel 1 in Fig.~\ref{fig2}), and reached a maximum size of $\sim$ 33\arcsec~around 13:00 UT. At this moment, the negative polarity peaked --1000~G, while the positive was about 800~G. Therefore, the magnetic field of both polarities increased from around +(-)200 to +(-)900~G in only six hours. A negative polarity on Sept 30, 11:24~UT, emerged and drifted to the right side with other negative polarity elements. These magnetic rearrangements are visible for the whole period and the loops in AIA 304~\AA~mark the polarity correspondence; however, no pronounced change in polarity is noticeable in HMI magnetograms, neither shear in AIA 304 nor 193~\AA~data. At the end of the sequence, the polarities diminished their flux and approached each other. The magnetic emergence seems to push the barb of the filament from below during the whole event. 

In the second row of panels in Fig.~\ref{fig2}, we plot AIA 211~\AA~images in mainly two lines: Fe~$\textsc{xiv}$, corresponding to $ log T$ $\sim$ 6.30; and Fe~$\textsc{xiii}$, corresponding to $ log T$ $\sim$ 6.25 \citep{ODwyer2010}. This wavelength happened to become very adequate to observe flux emergence patches under filaments, since the S/N is good, samples high temperatures, and the opacity of the filament becomes reduced but not totally suppressed. 

We also present evidence of an EUV counterpart of the flux emergence in 94~\AA~in the third row of panels in Fig.~\ref{fig2}. These images sample a temperature range in the corona \citep{ODwyer2010} with two lines: the inner quiet corona, sampled by Fe~$\textsc{x}$, corresponding to $ log T$ $\sim$ 6.05, and the flaring range, sampled by Fe~$\textsc{xviii}$, corresponding to $ log T$ $\sim$ 6.85.
The region started to increase its intensity from 6 DN to 30 DN. In \citet{ODwyer2010}, the quiet sun intensity is synthetically defined as 6 DN. Therefore, the intensity of this emerging region is five times that of the quiet sun, and 10\% of the maximum intensity in the flare of the full disk in this period.

This observational fact may suggest rapid connectivity between the photospheric longitudinal and transverse fields and the corona, since the signal starts to increase from 01:00 UT onwards, as shown in Fig.~\ref{fig2} and in the movie added as on-line material.

After the flare, we can see that opposite magnetic polarities rearrange and sometimes approach and disappear, displaying a brightening between them as a cancellation signature. In the post-flare panels, the last one is a very clear case.

Since we want to measure the magnetic flux emergence and check whether magnetic cancellation takes place, we compute the magnetic flux and magnetic flux density to this small area in HMI LOS-magnetograms. We set a threshold of +(-)90~G,  that is, $\approx$ 10$\sigma$ for 45-s magnetograms \citep{Liu2012}. This threshold is high enough to account for the flux computations in the inter-network and network, avoiding noise effects. Under all considerations, the filling factor is equal to 1.    

\begin{figure*}
\begin{center}
\includegraphics[scale=0.38]{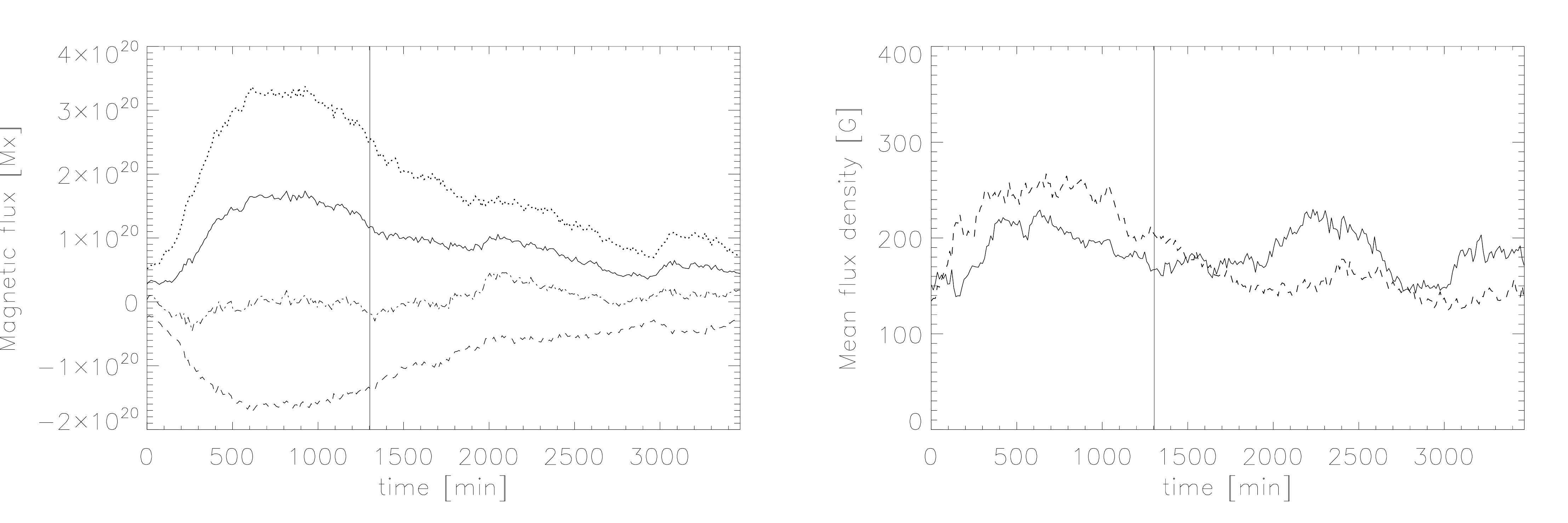} 
\caption{{\it Left:} computed magnetic flux on positive patches (solid line), on negative patches (dashed line), signed magnetic flux (dotted-dashed line), unsigned magnetic flux (dotted line). A vertical line marks the time of the flare at 21:43~UT. {\it Right:} mean values of the magnetic flux density of the positive (solid) and negative (dashed) regions.}
\label{fig3}
\end{center}
\end{figure*}

We  calculated the signed magnetic (positive and negative) flux separately, as  in the left panel of Fig.~\ref{fig3}. The magnetic flux has been corrected from heliocentric angle ($\mu$). 
 The vertical bars mark the flare onset. The starting time is Sept 29 at 00:00~UT. The positive flux reaches a maximum value at $\Delta t$=600~min, but there is a plateau from  $\Delta t$=600~min to $\Delta t$=960~min (from 10-16h~UT). Both positive and negative fluxes start to decrease around Sept 29 at 12:00~UT. The maximum unsigned flux is $\sim$3.4$\times$10$^{20}$ Mx, and the unsigned flux rate, 4.29$\pm$0.01 $\times$10$^{17}$~Mx min$^{-1}$. Calculating the emergence flux rate (up to the middle of the plateau) yields + 2.23$\pm$0.07 $\times$10$^{17}$~Mx min$^{-1}$. Considering the first maximum, the rate is +2.60$\pm$0.08 $\times$10$^{17}$~Mx min$^{-1}$.
We  performed the same operation for the negative polarity, obtaining --2.52$\pm$0.07 $\times$10$^{17}$~Mx min$^{-1}$.
For the negative polarity, we can see a clear decay in both magnetic flux and flux density, as shown in Fig.~\ref{fig3}. Calculating the decay rate of this negative area yields 5.2$\pm$1.2 $\times$10$^{16}$~Mx min$^{-1}$, which is much slower than the emergence rate.

The signed magnetic flux densities, shown in the right panel of Fig.~\ref{fig3}, are both around $\sim$+(-)250~G and follow the trend as the signed magnetic flux. There are two peaks on the positive magnetic flux density some time after the flare, probably caused by recurrent magnetic intensification in the area (see Section~\ref{s2}).

To rule out any possible contribution from other areas to the flux emergence in the region, we have also analysed three different areas where magnetic emergence is detected (close to the main sunspot of the active region [700\arcsec, 90\arcsec]. This area was actually flare-productive on Sept 29 at  05:11 UT, with the onset of a C1.6 flare. In the active region, the positive polarity decreased from 2$\times$10$^{21}$~Mx about 25\%, and the negative polarity increased from --1.5$\times$10$^{21}$~Mx by 16\%, becoming both fluxes equal 2.5~h before the flare. Furthermore, we studied the areas around coordinates [400\arcsec, 150\arcsec] and [400\arcsec, 30\arcsec], corresponding to supergranular voids, where magnetic flux emergence is also observed. For these regions and the threshold set, the flux emergence rate is negligible. 
We  computed the total magnetic flux of the whole AR as well, from September 24 00:00 UT to Sept 29 00:00 UT. The positive flux started at 0.9$\times$10$^{22}$~Mx, peaking on Sept 26 with 1.3$\times$10$^{22}$~Mx.  The negative flux  started at --1.1$\times$10$^{22}$~Mx, peaking at the same time. At the time of the flare, the fluxes were similar to Sept 24, keeping the decreasing trend. The rate of the total flux emergence respect to the AR is around 1\%. 

We also measured the expansion velocity of the emergence, with two main opposite polarities, from Sept 29~01:00~UT to 13:00~UT with different methods. The first method is the magnetic centroids, as detailed in \citet{Balmaceda2010}, but the distance was computed between the positive and negative polarities above the threshold of +(-)90 G. With this method, the expansion velocity is 0.165$\pm$0.006~\kms. The second method we used is visually measuring the distance between the two main polarities (as a diameter and obtaining the radius), getting 0.100$\pm$0.003 \kms. The third procedure is the one described in \citet{Palacios2012}, measuring the area and assuming it as circular, subsequently calculating the radius from it. This yields a velocity of 0.157$\pm$0.004 \kms.

\subsubsection{Linear polarization and transverse field of the flux emergence}

SDO/HMI retrieves a full-disk imaging of Stokes parameters, in the form of HMI full-disk Stokes spectrograms ({\it HMI.S{ \_}720s} data series). We  used Stokes {\it U} and {\it Q} images to create linear polarization images. These images are created with six points \citep[line sampling of about 69 m\AA~ from the centre line;][]{Schou2012, Centeno2011} in the Fe~$\textsc{i}$ line at 6173~\AA. Using the standard SSW routine for HMI/SP reduction, images are accumulated and processed. Profiles are extracted from the images and normalized by the red-most wavelength point in Stokes I ($I_{6}$).
For each frame, we have computed the linear polarization using the following formula \citep[e.g. ][]{Solanki2010, Pillet2011}:

\begin{align}
L_{pol}=\frac{1}{6}\sum^{6}_{i=1}\frac{ \sqrt{U_{i}^{2}+Q_{i}^{2}}}{ I_{6}}
.\end{align}

 In this stage, we also used also HMI inverted data by the Milne-Eddington inversion code VFISV \citep{Borrero2011, Centeno2014}. These data consist of the magnetic field, azimuth, and inclination to compute and plot the transverse field through $B\cdot sin(\gamma)$ with $\gamma$ the inclination with respect to the LOS, as described in \citet{Hoeksema2014}. For a general comparison between the LOS-magnetograms and the LOS-magnetic field obtained by weak field approximation through the Stokes profiles in active regions, see \citet{Palacios2015}.

\begin{figure*}
\begin{center}
\includegraphics[scale=0.415]{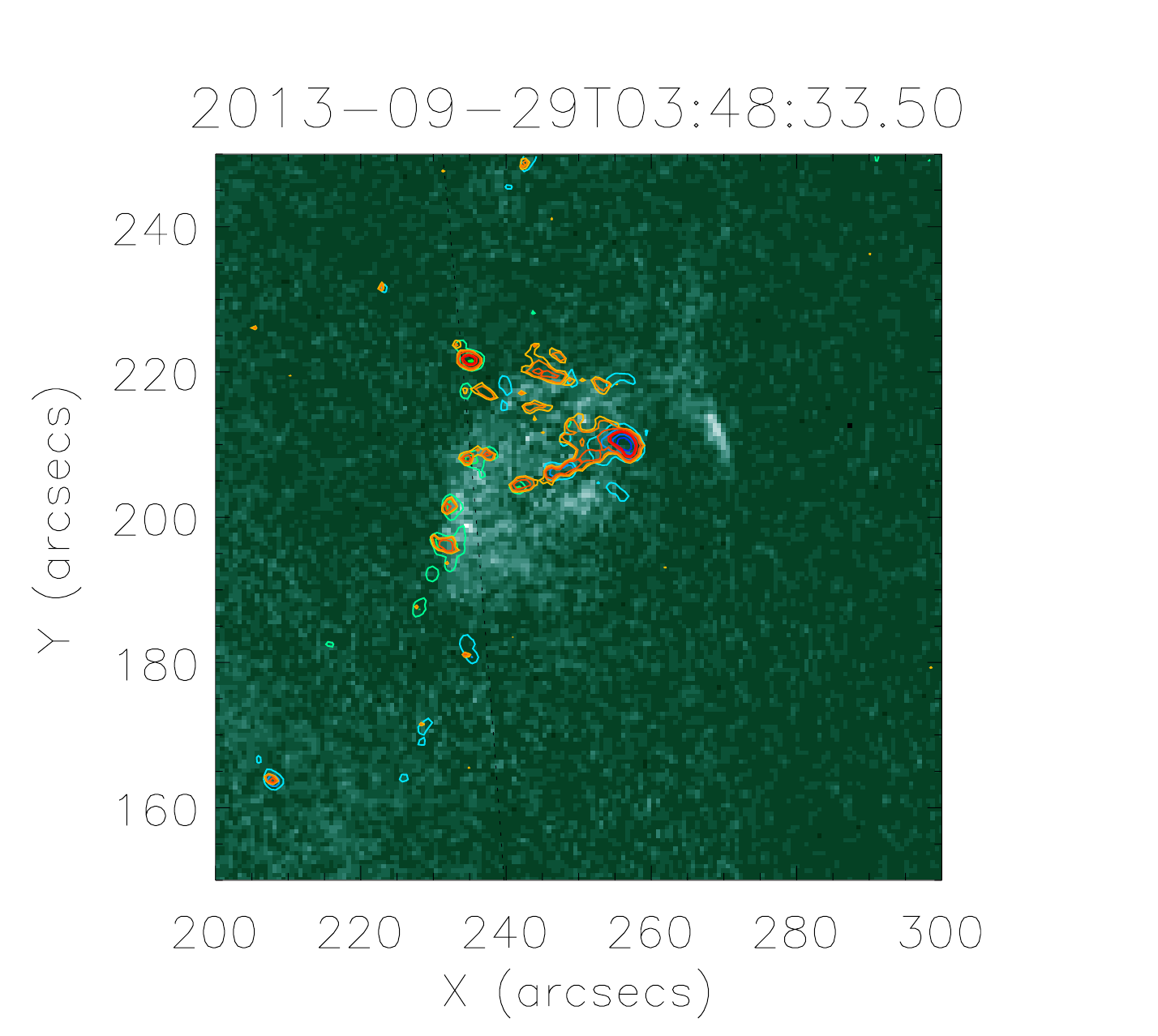} 
\includegraphics[scale=0.415]{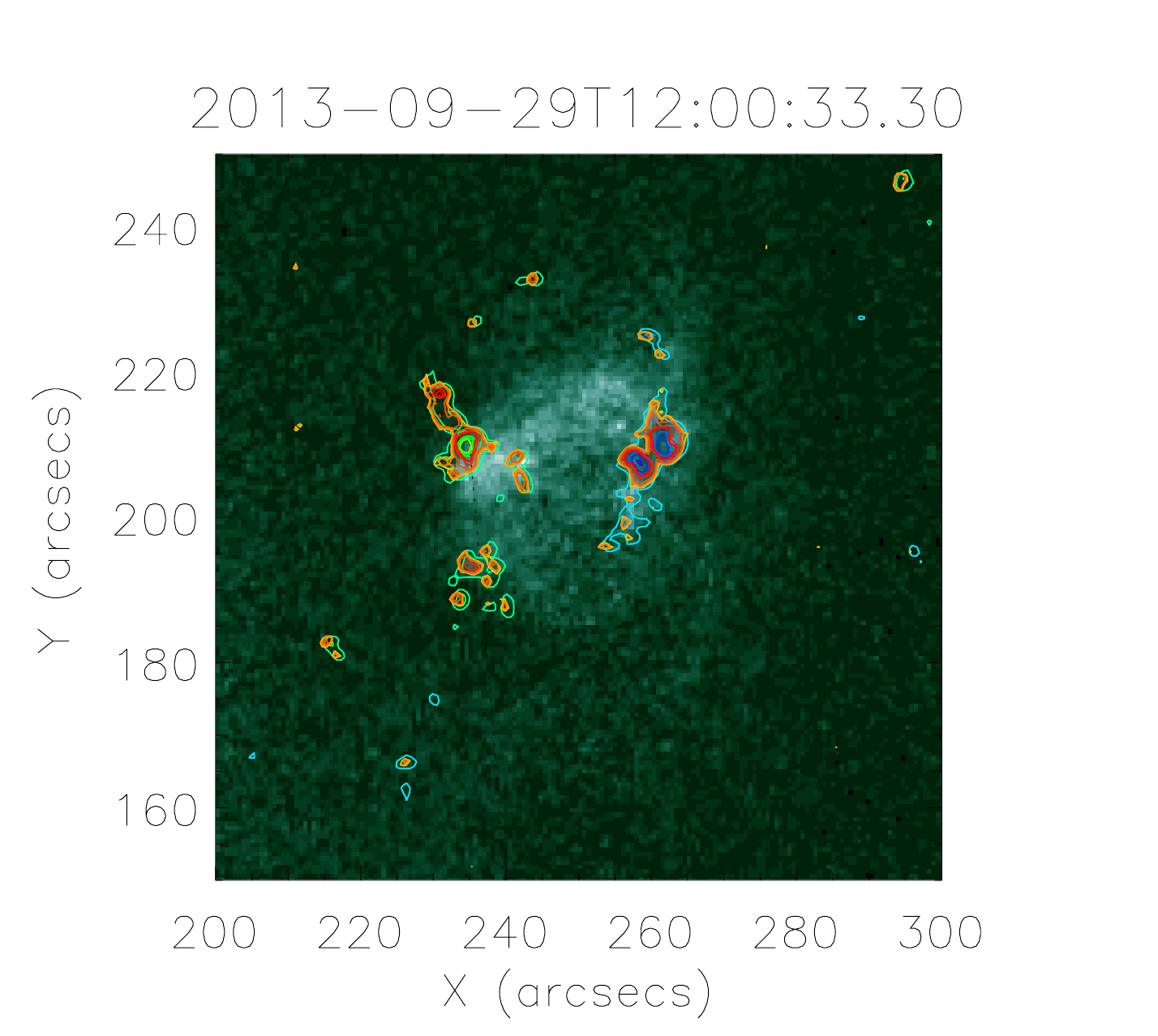}
\includegraphics[scale=0.415]{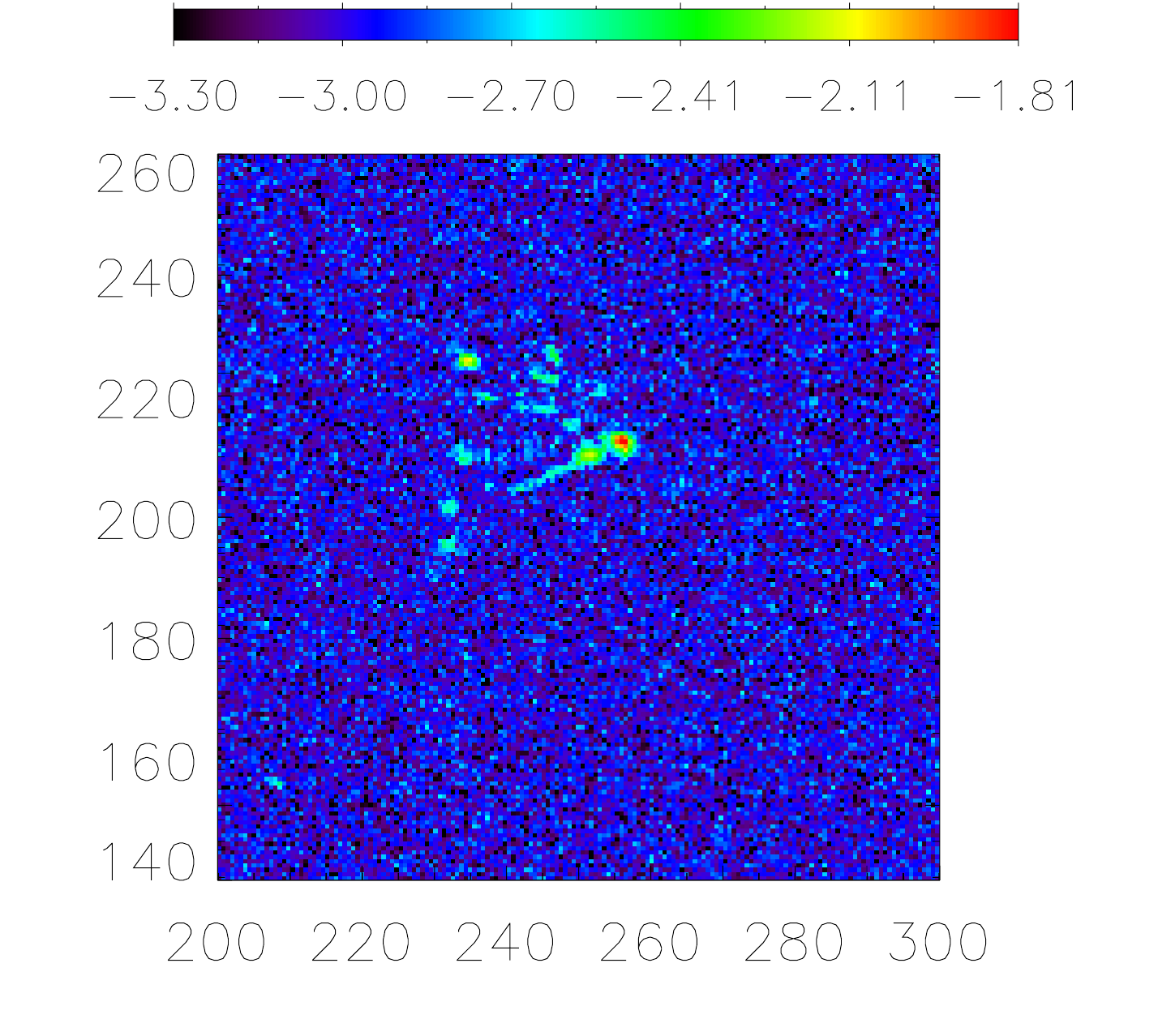}

\caption{{\it Left and centre:} AIA 94~\AA~image at different instants with LOS-magnetic field superimposed, with the same contours as Fig.~\ref{fig2}. Orange and red contours mark levels of 180, 200, 250, 400, 500 G in transverse field. {\it Right:} linear polarization map at 03:48 UT, plot in logarithmic colour scale.}
\label{figpol}
\end{center}
\end{figure*}

We studied the transverse field magnetic flux density. This includes a threshold of 180 G, which is double  the longitudinal flux. The magnetic flux density is included in Fig.~\ref{figpol}. In the left and central panels, we show both longitudinal (green and blue) and transverse field (shades of orange), with a 94~\AA~background, in two different instants: 03:48 and 12:00 UT. The rightmost panel shows the linear polarization at 03:48 UT plotted in logarithmic colour scale. The maximum linear polarization is 1.5\%.
The structure of the transverse field in the blob is not uniform, but creates bridges between the main positive and negative polarities. In the first four hours, this situation is conspicuous. After that, the transverse field patches remain in the main polarities because some inclination of the mainly vertical  magnetic field. Some showers of cosmic rays appear at some moments, altering the most sensitive part of the polarization images, i.e. the linear polarization, translated into transverse magnetic  field and making it look noisy.

\subsection{Filament rising}

\begin{figure*}
\begin{center}
\includegraphics[scale=0.33]{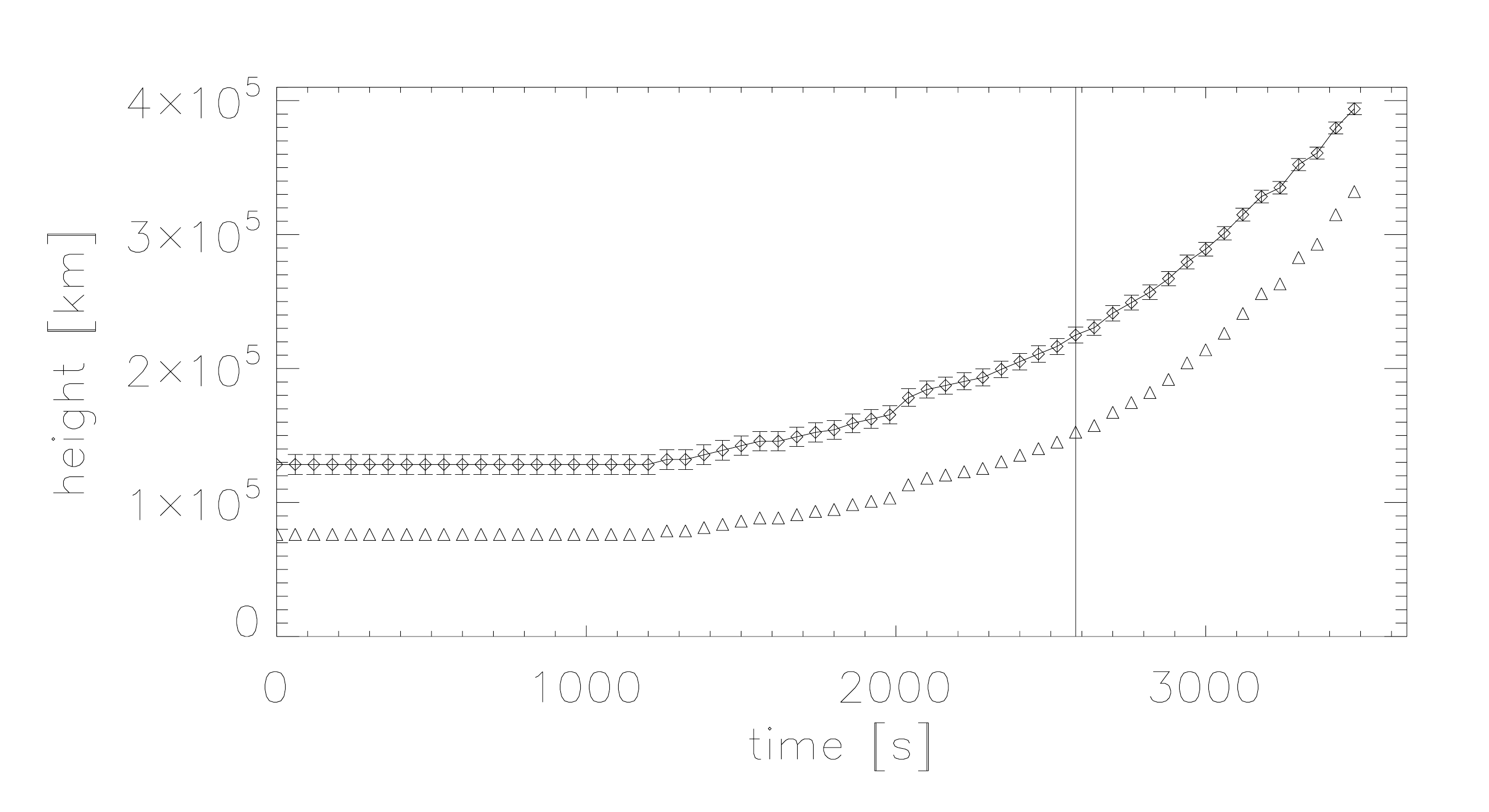}
\includegraphics[scale=0.33]{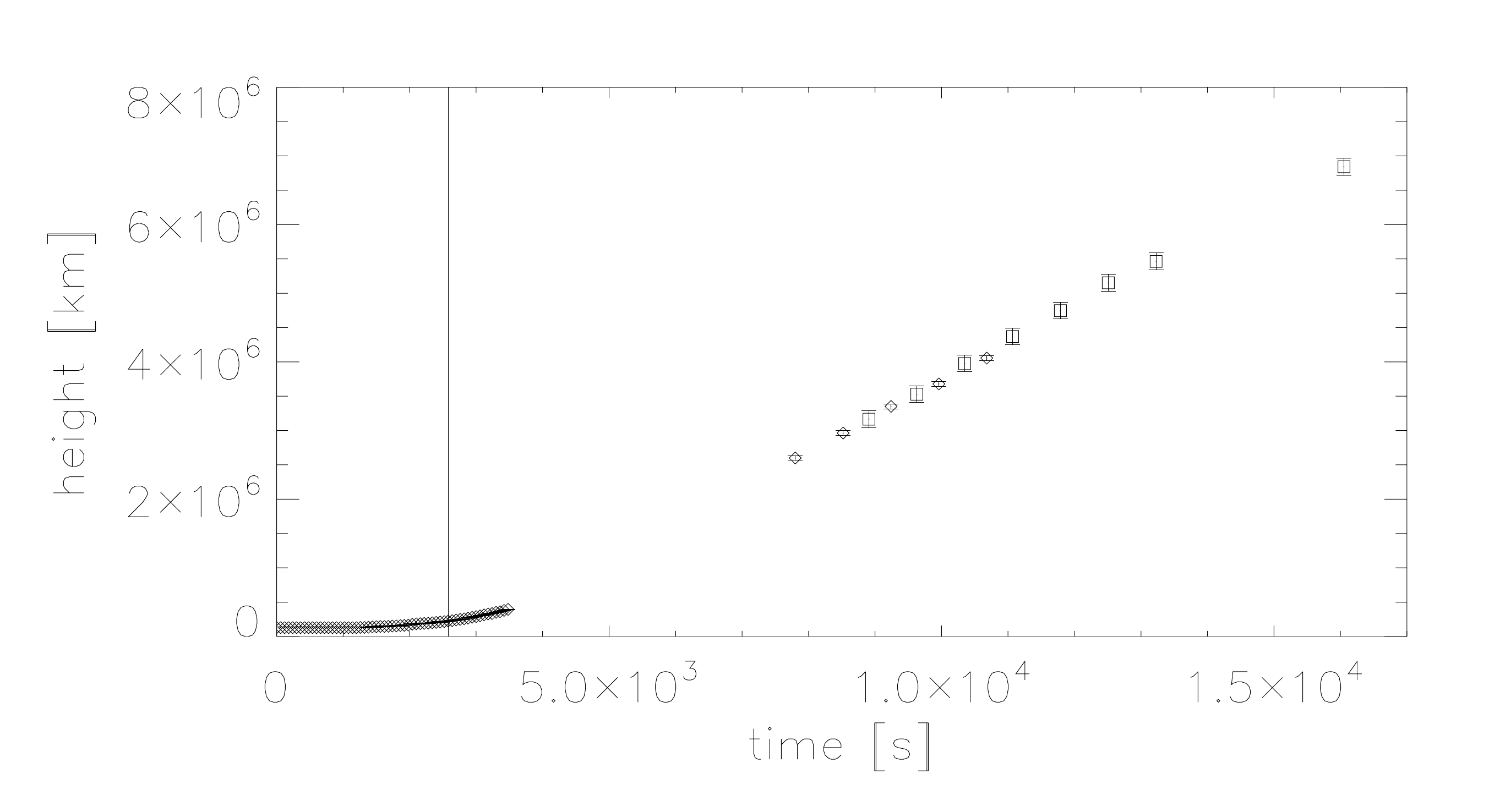}

\caption{{\it Left:} height versus time of the filament apex. Two plots are shown: triangles represent the uncorrected height data; and the solid line indicates the height after correcting of heliocentric angle projection. Error bars are computed considering the visual error on height and angle correction. Time starts from 21:00~UT and covers 1~h. In the last 38~min, the height starts increasing. A vertical bar marks  the flare onset, at 21:43~UT. {\it Right}: height-time diagram, including LASCO C2 (diamonds) and C3 (asterisks) data.}
\label{fig4}
\end{center}
\end{figure*}

We calculated the rising speed on the apex of the filament on de-rotated AIA 304~\AA~images, also provided  in the on-line material, and whose detailed reference frame is shown in the right panel of Fig.~\ref{fig1}. The cadence of this series is 1~min from 21:00 to 22:00~UT. We examined the previous frames from 20:00-21:00~UT, but the rising is negligible at that period. As shown in Fig.~\ref{fig4}, the filament  kept a constant height during 22~min, and the lift-off took around 38~min.The lift-off of the filament started much before the flare onset at 21:43~UT. The height was estimated on de-rotated images, considering a fiducial point [340\arcsec, 340\arcsec] that lies in the line of the axis defined by the two-ribbon flare middle point and the footpoints. The distance between footpoints is around 0.8 $R_{\sun}$.The apex height was measured, albeit the southern part of the filament was more active on rising the first 20~min. These measurements are indicated with triangles in Fig.~\ref{fig4}. The visual error for the uncorrected height is considered as 3 pixels of a rebinned image of 1024 $\times$ 1024 pixels, translated as 5220~km (not shown). We corrected from projection of the heliocentric angle, i.e. dividing by the heliocentric angle sine, considering the eruption follows the local surface normal direction, as shown in the plots of  Fig.~\ref{fig4}. The mean velocity from the flare to the end at 22:00~UT is 188$\pm$5~\kms, and the subsequent acceleration in this period is 0.065$\pm$0.007~\kmss. The maximum velocity is 312~\kms. (Considering the 38-min period that the rising lasts, the velocity is 115$\pm$5~\kms~ and acceleration is 0.049$\pm$0.001~\kmss~).

For this event, we fit the ascent height to different functions, since this can shed light on the physical mechanism that triggered the eruption. We tried exponential,  $\it{log-log}$, and parabolic function. Then we evaluated them with the merit function $\chi^{2}$ (unreduced). The exponential fit for the time-height (uncorrected) yields $\chi^{2}$=1.7, while the $\it{log-log}$ fitting yields $\chi^{2}$=5.9. In the case of projection-corrected height, the fitting is even better: $\chi^{2}$=0.9, while the $\it{log-log}$ fitting yields $\chi^{2}$=3.4. Parabolic fittings get very large $\chi^{2}$.

We also  completed the study with SOHO \citep{Domingo1995} LASCO C2 and C3 level 1 data \citep{Brueckner1995}. After centring and rotating,  we  performed image base differences and then, applied a Sobel operator  for enhancement. We  measured the apex in the filament images. The plate scale for C2 is 11.9\arcsec pix$^{-1}$ and for C3 is 56\arcsec pix$^{-1}$. We show the whole time-height scale for AIA304, C2 and C3 in Fig~\ref{fig4} (bottom). The $\chi^{2}$ for the whole sequence is 3.80.

 \subsection{Potential field modelling and critical decay index for the filament region and emergence}\label{pfsss}

\begin{figure*}
\begin{center}
\includegraphics[scale=0.30]{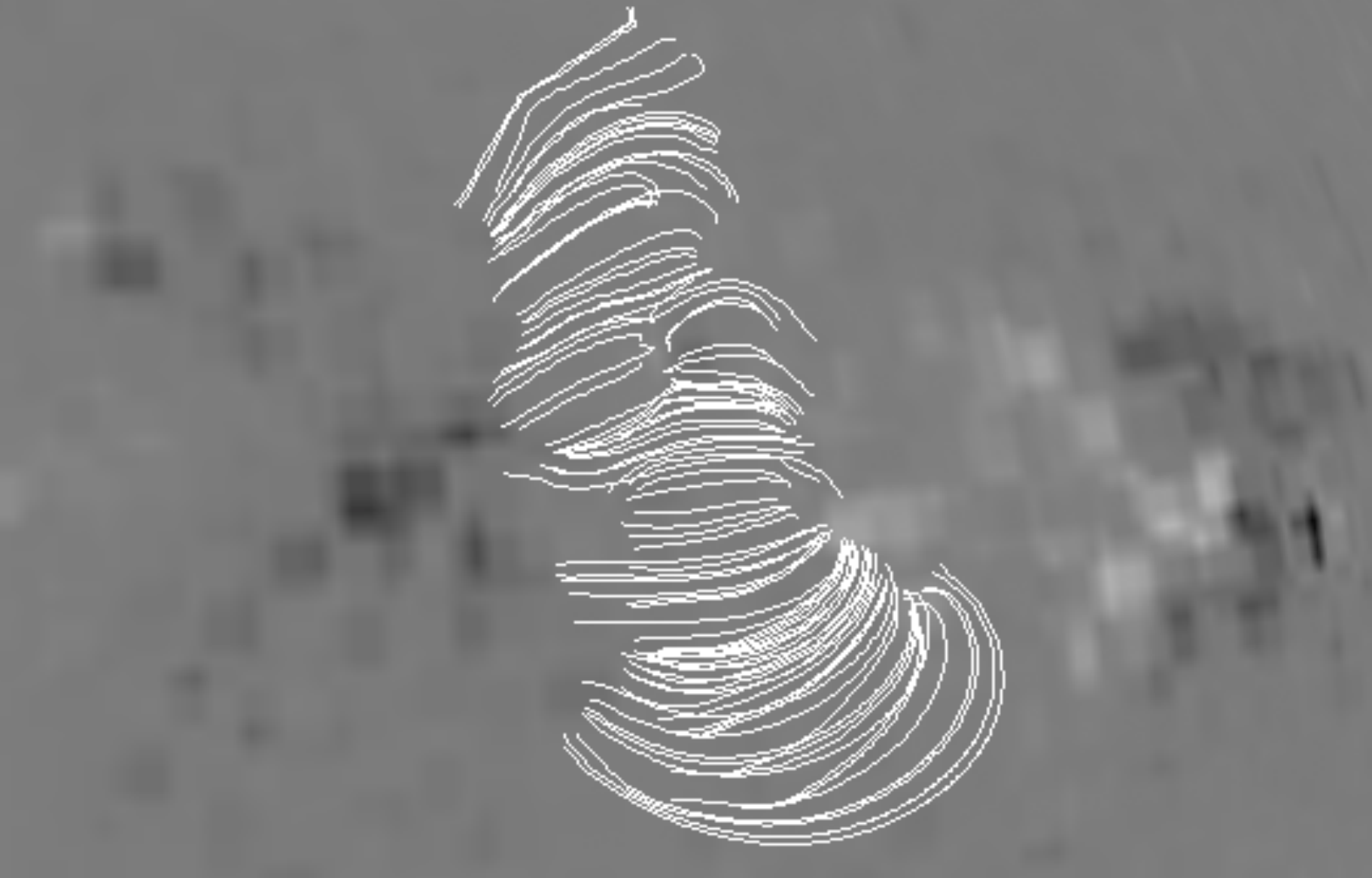} \includegraphics[scale=0.182]{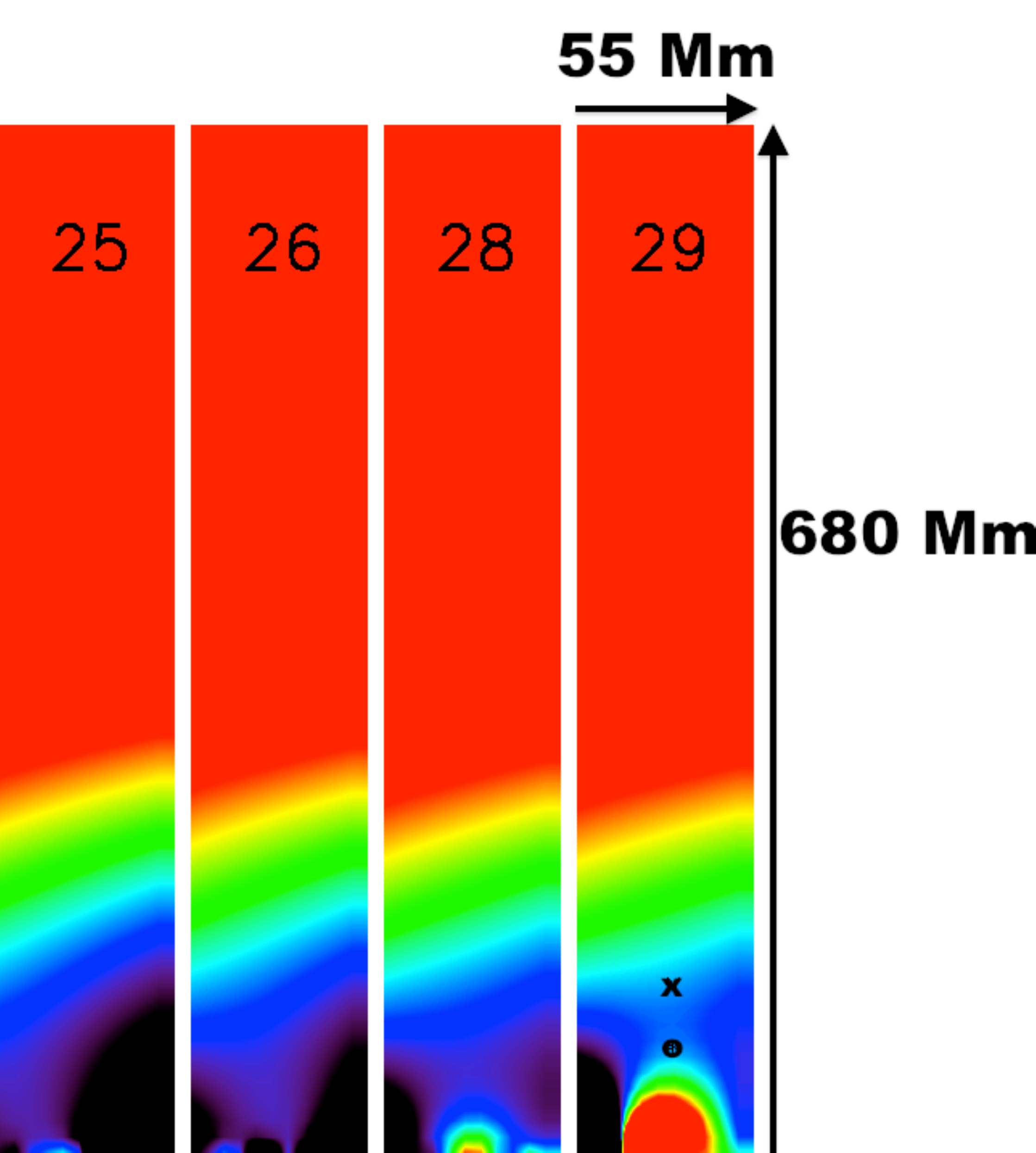} 
\includegraphics[scale=0.23]{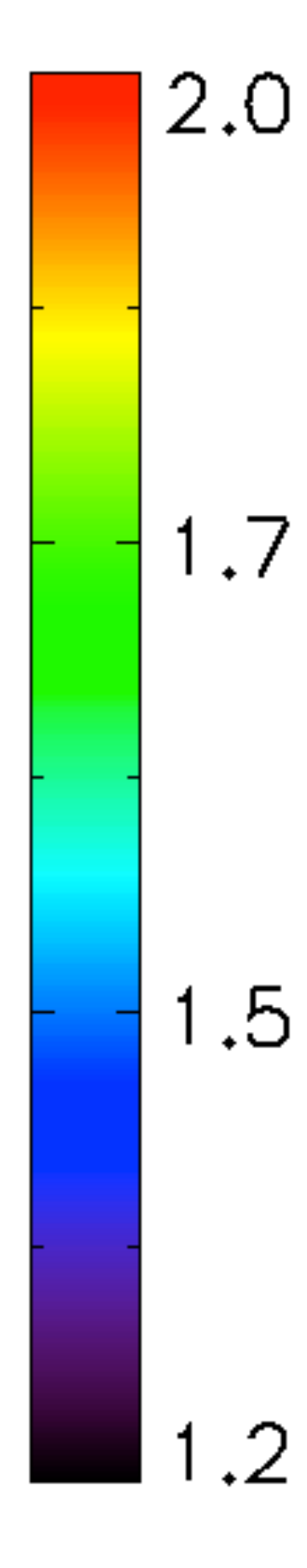} 
\includegraphics[scale=0.355]{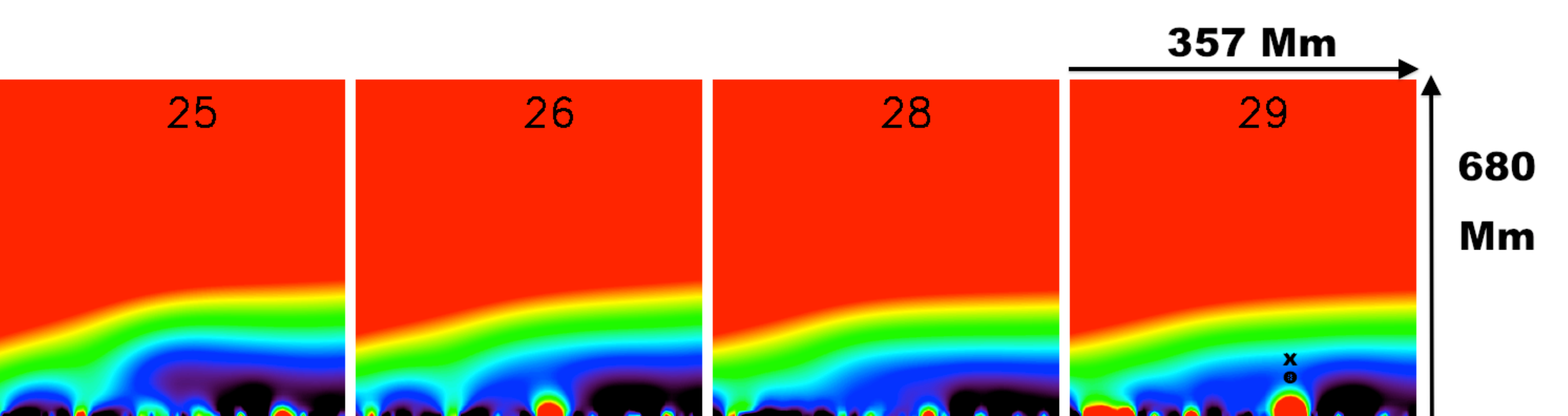} 
\caption{{\it Top left:} PFSS magnetic rendering corresponding to the filament region rendered by {\it PFSS} for Sept 29 data. A low-resolution magnetogram was used, with the same FOV as Fig.~\ref{fig1} (right). {\it Top right:} transverse slice of the decay index $n$ across the volume that encloses the filament and emergence. The estimated height of the filament for Sept 29 is marked with a black cross (corrected), or small circumference (uncorrected).  {\it Bottom:} longitudinal slice of the decay index $n$ along the volume that encloses the filament and emergence, and same symbols for the filament height. This figure belongs to subsection~\ref{pfsss}.}
\label{pfss_cuts}
\end{center}
\end{figure*}

We  used the package Potential Field Source Surface available in SSW {\it PFSS} \citep{Schrijver2003} to assess the critical height of the filament through the decay index of the external field $n$, since we aim to know whether the filament was already unstable before the emergence or the emergence made the filament unstable. The underlying model for the decay index $n$ is the torus instability. The input of the {\it PFSS} model is a synoptic magnetogram. We  used the highest resolution input data,  the available daily quasi-synoptic HMI magnetograms, obtained at 12:00 UT, on Sept 25, 26, 28, and 29 (27 not available). These synoptic maps were not derotated, so care had to be taken to identify the areas. We  set the {\it PFSS} computation grid as 800 points in solar longitude, 400 in latitude, and 66 points in height, covering from 1$R_{\sun}$ to 2$R_{\sun}$. Therefore, the spatial sampling of the grid is 0.45\deg~per grid point in longitude and latitude, and around 10600 km per grid point on height. We included a context image of  the {\it PFSS} loop set for this region on Sept 29, 18:00 UT (top panel of  Fig.~\ref{pfss_cuts}). The height rendered is in agreement with the estimated filament height.

Then, with the decay index of the external field $n$, used recently in \citet[][ and references therein]{Kliem2006, Torok2007, Zuccarello2014}, we  estimated this magnitude as 

\begin{align}
n=-R \frac{\partial{ln(B_{ex})}}{\partial{R}} 
.\end{align}

We  generated datacubes of $n$ for Sept 25, 26, 28, and 29. We  selected a box of 10$\times$65$\times$66 grid points, equivalent to 4\deg5 $\times$29\deg25 $\times$1$R_{\sun}$ (solar longitude, latitude, radial height). The $Z$ axis actually goes from $\sim$ 20 Mm to 700 Mm, since we removed the  points on the very bottom. This final volume of $\sim$55~Mm $\times$ 357~Mm $\times$ 680~Mm includes the filament and  flux emergence. The $Y$ axis covers approximately the filament's length, but considering its shape, it cannot be considered as wholly lying in the solar longitude selected plane. We clipped the $n$ values  from 1.2 to 2.0. We cut this box along and across the exact location where the flux emergence happens, as shown in Fig.~\ref{pfss_cuts}. 

The decay index $n$ is represented in shades of purple-blue on the volume where it is torus-unstable  \citep[$n$ becomes a critical factor for instability, $n_{crit}$, when values  ranges from 1.3 to 1.5, as in][]{Zuccarello2014}. Considering the low impulsivity of the ejection, this threshold is also in agreement with  \citet{Torok2007}.

During the previous days, some network points are rooted in the area, but the most important variation are from Sept 28 12:00 UT to Sept 29 12:00 UT. The height of the filament may vary and is only marked for Sept 29, with a hollow circle the projection-uncorrected height, and the corrected height with a cross. On Sept 28, $n$ increased on the emergence area. Black areas are those  where $n<$1.2, and these areas decreased from 25 onwards. On Sept 26, a large network area also made $n$ increase, but it was not located in same coordinates of the emergence. 

\section{Discussion}\label{s2}

The emergence of this relatively small-scale region follows Hale-Nicholson's law of polarities \citep[the same leading and trailing polarity in the same hemisphere, ][]{Hale1919}; in this case,  negative is leading, and positive is trailing, as the leading AR11850. However, in the right panel of Fig.~\ref{fig1}, we  notice a different magnetic configuration. The filament is located between the negative polarity of the next active region (to the left) and the positive polarity from AR11850 between where we can locate the polarity inversion line (PIL). Therefore, the emergence exhibits a normal configuration, but inverse to the filament configuration, which can help the magnetic instability. The situation is similar to the simulations that originate eruptions in \citet{Chen2000}, as an opposite polarity configuration emerges beneath or at the side of a filament. Observing in different coronal wavelengths, as 193 \AA~and 94~\AA, there is no evidence of a large magnetic canopy over the filament in the form of bright loops. LASCO images also reveal a leading edge and the filament inside this edge. These images seem to be in agreement with the standard model of a twisted flux rope surrounded by a less twisted longitudinal magnetic field, but without including a large-scale magnetic arcade. The inverse-S shape and  barbs would define it as dextral chirality, and the longitudinal field of the filament would then point to the solar north direction. The twisted field lines would exhibit negative helicity.

We have to note that this filament configuration (dextral)  follows the hemispheric pattern; however, exceptions to this rule are reported \citep[e.g. see][and references therein]{Mackay2014}. Furthermore, under a full polarimetry analysis, the chirality definition may change, from a topological orientation of features to a real estimation of the magnetic field direction and sense \citep{Hanaoka2014}. 

The barb is very close to the positive polarity from before the supergranular flux emergence, and seems  fundamental to the mass dynamics of the filament. Since the barb is very close to the emergence, it may be  the point where the instability is transmitted. 

The rate of flux emergence, which is about 2$\times$10$^{17}$~Mx min$^{-1}$, is at least one order of magnitude larger than those reported by \citet{Palacios2012}. However, the expansion velocity of the flux emergence is much slower,  at 0.16~\kms~ , than in Palacios et al., which is approximately 0.6~\kms~. In this case, the higher magnetic fields  may slow down the velocity of the upflowing plasma. 
The expansion velocity calculated with centroids is similar.  
 
Expecting dramatic changes in photospheric magnetic flux as large reconnection features, we only found the bright blob in 94~\AA, being conspicuously bright in 211~\AA. The magnetic flux emergence lasts around 12 hours and smoothly decays, probably due to the reconnection that started a bit after the flux emergence detection. We  included spectropolarimetric data to clarify the magnetic structure by adding the transverse component during the whole emergence. We may expect a larger transverse field structure, but that may be smeared out because of the averaging over 12 minutes, or some circular to linear cross-talk. Combined images between AIA 94~\AA ~images and longitudinal magnetograms suggest  the magnetic link between the emergence in the photosphere and the coronal counterpart, in the form of a blob, brighter than the surroundings. This coronal counterpart is also visible in 304 and 193~\AA, but somewhat occulted by the barb due to its opacity. The blob is best seen in 211~\AA\  since the emergence patch is better seen through the filament barb, as the latter is less opaque. 

Compared to simulations of \citet{Kusano2012} and \citet{Toriumi2013}, this emergence region resembles an OP-type configuration with azimuth angle $\phi_{e} \sim$ 160\deg, and shear $\theta_{0}$ $\sim$ 60\deg ($\phi_{e}$ is the angle between the PIL and the orientation of the bipolar region. With this shear and according to these simulations, the injected kinetic energy might be large enough to provoke an eruption. \citet{Bamba2013} states that OP-type may influence the destabilization of a flux rope via reconnection.

We followed the evolution of the filament for about six days, and the phenomenon known as the "sliding door effect" \citep{Okamoto2008, Okamoto2009, Kuckein2012} is not evident, since the distance between the positive and negative facular boundaries  remains the same, $\sim$220\arcsec. The filament was already there  at the beginning of the study, and the part that grew from there does not exhibit any distinctive pattern of the sliding door effect.

Development of pores have been reported by \citet{Kuckein2012}, and this is considered  tracers of the filament emergence. However, \citet{Vargas2012} had studied the event of  \citet[][]{Okamoto2008, Okamoto2009} and considered these photospheric signatures as necessary but not sufficient pieces of evidence for supporting the hypothesis of emergence. We also find very small  appearance of pores. However, it is probably more related to the intensification of the magnetic field \citep{Grossmann1998, Bellot2001, Danilovic2010} rather than
characteristics of another flux rope emerging, since the magnetic field density increases from -200 to -1000 G . In the right panel of Fig.~\ref{fig3}, other peaks appear in the positive magnetic field density, probably caused by recurrent magnetic field intensification in the region.

Another point is that the CME speed was reported as 600 \kms. In the case of CME velocities, we always should have in mind the solar escape velocity, that is 617 \kms. Any difference from this velocity can account for the projection angle or the kinematic energy injection to the CME.

Furthermore, we  checked  that CH1 becomes darker visually, which might be an indicator of more open field lines there. Some of these features will be treated in another paper.

Regarding the filament eruption, the current instability models might explain or rule out the mechanism that causes the eruption. In the review of, e.g. \citet{Aulanier2014}, different models are presented. The breakout model \citep{Antiochos1998, Antiochos1999} depends on significant shearing, and involving reconnection above the flux rope \citep{Schmieder2013}. Reconnection is strongly enclosed in some of these models \citep{Zuccarello2013}. Loss of equilibrium models \citep{Forbes1991, Forbes1995} may explain the slow filament rising \citep{Schmieder2013}. These models rely on photospheric motions, polarity approaching, or vortical motion \citep{Amari2003, Aulanier2010}. Another model to consider is "tether-weakening", with observational examples  presented by \citet{Sterling2005, Sterling2007}. This weakening is usually due to emergence off the filament channel \citep{Moore1992}.

In addition to the observational signatures of the models, the height-time fitting may provide hints about the physical mechanism. In \citet{Schrijver2008}, the expansion is fit with different functional forms (polynomial, exponential, power-law...) and evaluated with the merit function $\chi^{2}$. Time-height exponential fitting is also used in very rapid CMEs, like in the case of \citet{Gallagher2003}. \citet{Gosain2012} also uses fitting to exponential. The time-height fittings presented here gets better results for exponential functions than any other, and even better to projection-corrected height than uncorrected. Actually, an exponential ascent may be a characteristic of different ideal instabilities, e.g. "torus instability" at the first stages \citep{Schrijver2008, Gosain2012}. Observing the filament expansion and its shape at several solar radii, it may reveal a kink instability, as in \citet{Torok2005}.

We checked the decay index $n$ of the area from Sept 25 to 29, similar to the study by  \citet{Zuccarello2014}.\ The parameter $n_{crit}$ is always between 1.3 and 1.5 on the filament and emergence volume, which may indicate a torus instability. However, the lower part of the volume exhibits an important change of $n$ in the area corresponding to the flux emergence, proving useful to identify ongoing instabilities. Importantly, $n_{crit}$ ranging the mentioned values may also imply an exponential-to-linear expansion \citep{Kliem2006}.

After exploring the main models of eruption in the above paragraphs, ideal instabilities (torus instability and also kink) can be the most plausible mechanism. We may establish a possible scenario as follows: the filament is destabilized, and a large magnetic flux emergence below the filament spine can help the destabilization, elevating against gravity and magnetic tension, increasing the magnetic pressure. 

The filament may be located close by a neutral point, which might be metastable (the paler shade of blue in Fig.~\ref{pfss_cuts}, which is the minimum corresponding $n$ in the $Z$ direction). This point may allow the topology of the magnetic field to be maintained without 3-D reconnection with the lower part of the filament, and may  help the torus instability to develop due to the flux emergence beneath this point. These kind of eruptions may be slow, leading to CMEs, and with one/several flux emergence patches underneath. 
 
\section{Summary and conclusions}\label{s3}

This filament, located in the boundaries of the active region AR11850,  evolves during almost five days of observations and eventually  erupts. On Sept 29, 00:00 UT, a supergranular flux area started emerging and about 20~h later, the filament started rising. The most conspicuous rising phase lasted about 38 min and the C1.2 flare onset started at 21:43 UT, while the two-ribbon arcade started developing around 15 min later. 

The height-time profile corresponds to an exponential profile, and the theoretical model that fits this smooth ascent most is the torus instability, plausibly related to a large magnetic flux emergence, and supported by the height-time profile and the succession of events (first the filament eruption and the flare afterwards). 

\section*{Acknowledgements}
The authors want to acknowledge SDO/AIA and SDO/HMI Data Science Centers and Teams, and the GOES and LASCO teams. The LASCO catalog is a CME catalogue that is generated and maintained at the CDAW Data Center by NASA and The Catholic University of America in cooperation with the Naval Research Laboratory. SOHO is a project of international cooperation between ESA and NASA.
We would like to thank to the Virtual Solar Observatory (VSO) and Helioviewer for data acquisition and visualisation, respectively. Also, we acknowledge data use from WDC from Geomagnetism, Kyoto, and LMSAL Solarsoft. This research has made use of NASA's Astrophysics Data System. We would like to thank funding from the Spanish project PPII10-0183-7802 from ``Junta de Comunidades de Castilla -- La Mancha'' and MINECO project code AYA2013-47735-P. We want to thank the anonymous referee for useful comments.

\bibliographystyle{aa} 

\bibliography{./aa_fil.bib}

\end{document}